\documentclass[preprint,showpacs,showkeys,aps]{revtex4-1}
\usepackage{units}
\usepackage{graphicx}
\usepackage{epsfig}
\usepackage{amssymb}

\begin{document}

\title[]{A many-body overview of low-energy optical excitations in armchair graphene nanoribbons}
\author{Jessica Alfonsi}
\affiliation{Universit\`{a} di Padova, Italy}
\email{Corresponding author: jessica.alfonsi@unipd.it} 

\begin{abstract}
Excitonic spectra of armchair graphene  nanoribbons (AGNRs) obtained from a full many-body exact diagonalization of the Hubbard model are reported for both longitudinally and transversely polarized photons, thus providing a complete survey of low-energy may-body optical excitations in these systems. 
The resulting one-photon allowed eigenstates turn out to be well separated in energy from each other but both couple to the same set of two-photon allowed states. The magnitude of the calculated optical oscillator strengths for perpendicular polarization suggest that these optical features can be indeed observed in polarized absorption measurements.
\end{abstract}



\maketitle

\section{Introduction}
The recent progress in the synthesis and isolation of single graphene-layers \cite{Geim2007,Wu2010,Cooper2012} has considerably increased the interest in graphene nanoribbons (GNRs), infinitely long stripes of carbon atoms arranged in a honeycomb lattice. The possibility of tuning the width of these quasi-one-dimensional (1D) nanostructures alongside their  edge geometry provides a strategy to overcome the absence of an electronic gap in graphene, which is one of the main limits preventing the application of this material in electronic devices \cite{Chen2007,Han2007}. Recently, high quality stripes less than $\unit[10]{nm}$ wide, which can be exploited for making field-effect transistors, have been indeed fabricated from bottom-up techniques \cite{Kosynkin2009,Jiao2009,Jiao2010,Cai2010}, thereby opening the path for a range of experimental investigations on the optical and transport properties of these systems. This will be a crucial benchmark for testing the theoretical predictions provided until now for GNRs, which have all been recently surveyed in several excellent reviews \cite{CastroNeto2009,Abergel2010,Marconcini2011,Wakabayashi2010}. Most of these studies rely on independent-particle approximation with proper boundary conditions at the GNR edges, such as tight-binding (TB) methods for $\pi$ electrons \cite{Ezawa2006}, the $k \cdot p$ two-dimensional Weyl-Dirac equation for free massless particles whose Fermi velocity $\left(\unit[10^{6}]{m/s}\right)$ plays the role of an effective speed of light \cite{Brey2006} and \textit{ab initio} density functional calculations \cite{Son2006}.
However, since electron-electron interactions are expected to be enhanced in such low dimensional systems, there have been also several theoretical investigations  concerning the inclusion of many-body effects beyond single-particle theory. These rely on \textit{ab initio} GW-Bethe Salpeter methods\cite{Louie2007,Prezzi2008}, the solution of Pariser-Parr-Pople effective model Hamiltonians for $\pi$ electrons incorporating longer-range Coulomb interactions \cite{Shukla2011,Mazumdar2012} and Hubbard model based approaches, either within mean-field approximation \cite{Fujita1996} or beyond it with configuration interaction (CI) carried out at different levels \cite{Pati2008,Alfonsi2012}. The suitability of mean-field approximation for the Hubbard model when applied to graphene-based systems has been matter of recent debate in literature \cite{Cini2008,Honecker2010,Yazyev2010,Palacios2010}, in view of the low-intermediate value for the Hubbard correlation coupling strength $\left( 1 < U/t < 2.2 \right)$ which seems plausible for these systems.
The geometrical classification of GNRs is dictated by their edge shape, which can be zigzag $\left(Z\right)$, armchair $\left(A\right)$ or chiral  $\left(C\right)$. Moreover, the structure of armchair nanoribbons (AGNRs) can be derived from that of zigzag single-walled carbon nanotubes (ZSWCNTs) by unzipping the graphene cylinder along the nanotube axis, while conversely zigzag GNRs (ZGNRs) can be obtained from armchair SWNTs. This concept has been actually exploited for producing controlled-sized GNRs through chemical \cite{Kosynkin2009,Xie2011} and laser \cite{Kumar2011} unzipping of multi-walled CNTs and it has also been recently extended to the fabrication of graphene quantum dots starting from fullerene cages \cite{Lu2011}.\\
The absence of zero-energy localized states simplifies considerably  the investigation of the electronic properties of AGNRs in comparison with ZGNRs. Besides that, the electronic band picture provided for AGNRs closely resembles that of ZSWCNTs, since the van Hove singularities occur at the centre of the Brillouin zone \cite{Fujita1996,Zheng2007,Wakabayashi2011}.
On the same footing, the optical selection rule for AGNRs with light polarization parallel to the GNR edge is clearly reminiscent of that found for conserved-quantum number interband transitions in ZSWCNTs, as discussed in several recent works reporting on optical selection rules and analytical expressions for the TB electron-light interaction matrix elements in both armchair and zigzag GNRs with either longitudinally or transversely polarized photons \cite{Sasaki2011,Lin2011,Sasakiunpub,Hsu2007}.\\
Both TB and Weyl-Dirac equation predict that armchair GNRs with pristine edges may be either semiconducting or metallic according to their width (oscillating gap). However, \textit{ab initio}  and mean-field Hubbard model results, recalled in the reviews by Cresti \cite{Cresti2008} and Rozhkov \cite{Rozhkov2011}, point out an always semiconducting behaviour for AGNRs, since the metallic state is unstable against bond deformations at the edges, electron-electron interactions and longer-range hoppings.\\
This work expands our previous full-many body exact diagonalization (ED) results for the Hubbard model applied to investigate optical excitations in AGNRs with longitudinally polarized photons \cite{Alfonsi2012} by considering also the polarization component perpendicular to the GNR edge, thus providing a complete overview of excitonic effects in these systems. Two-leg ladder models mimicking AGNRs of different widths and belonging to three distinct families are used to sample the set of k-points lying at the centre of the Brillouin zone which give the van Hove singularities and the maximum absolute values of the interband optical matrix elements. The appropriate velocity operator for optical transition is mapped onto these quantum lattice models to give the chosen polarization component in the calculation of one- and two-photon absorption optical matrix elements.  The obtained trends for the allowed transitions and related optical oscillator strength as a function of the correlation coupling strength $U/t$ provide an insight into the family dependence of the optical anisotropy in AGNRs and suggest that the spectral features related to transversely polarized photons can be observed in these systems in polarized absorption measurements.

\section{Method}
\begin{figure}[htbp!]
\epsfxsize=8cm
\epsfysize=7cm
\centering
\includegraphics[scale=0.7]{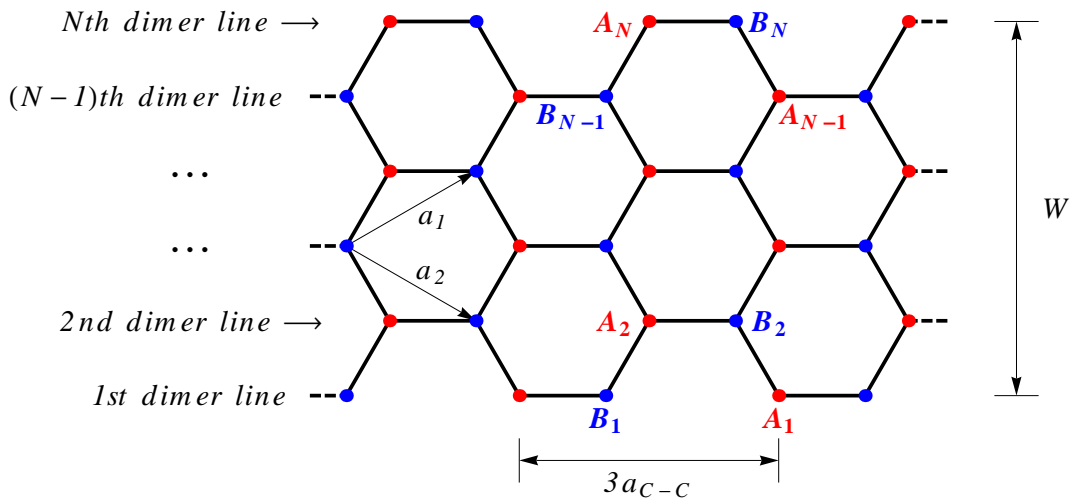}
\caption{(color online) Armchair nanoribbon of width \textit{W} with N dimer lines.  \textit{A}(\textit{B})-type atoms are shown in red (blue), respectively. Equivalent atoms on the honeycomb lattice are mapped by the $a_{1}$ and $a_{2}$ basis vectors.}
\label{agnrlattice}
\end{figure}

In Fig.\ref{agnrlattice} we show the geometrical structure of an armchair graphene nanoribbon unit cell. Its width \textit{W} is equal to the number \textit{N} of dimer lines containing A-type and  B-type carbon atoms, hence a total number of $2N$ sites. Periodic boundary conditions along the direction parallel to the edges are represented through dashed lines. Following the traditional nomenclature for GNRs \cite{Son2006}, we consider in this work AGNRs with $4 \leq N \leq 7$.  Incidentally, we note that such small-width structures are not merely hypothetical, since the 7-AGNR structure has recently been obtained from  aromatic precursors and  investigated both by scanning tunneling microscopy and Raman spectroscopy \cite{Cai2010}.
\\
Since \textit{N}-AGNRs are topologically equivalent to to brick type lattices (i.e. with periodic ladders) \cite{Son2006,Cresti2008} and periodic boundary conditions along the direction parallel to the nanoribbon edges are considered ($k_{||}=0$ states are sampled in the GNR Brillouin zone), one can fold the brick type lattice into a two-leg open ladder having \textit{N} rungs and a unique value \textit{t} for all nearest neighbour hopping parameters in the Hamiltonian (see Fig. 3 in Ref. \cite{Son2006}).
We consider the simple Hubbard Hamiltonian form for $\pi$ electrons
\begin{equation}\label{hamhubbard}
H^{2D}=-t_{\pi} \sum_{\langle i,j\rangle , \sigma}\left( c_{i,\sigma}^{\dagger}c_{j,\sigma} + h.c.\right)+U\sum_{i}n_{i,\uparrow}n_{i,\downarrow}
\end{equation}
where \textit{i} and \textit{j} are site indices, $\langle i,j \rangle$ are all pairs of first nearest neighbor sites, $c_{i,\sigma}^{\dagger}$ and $c^{•}_{i,\sigma}$ are electron creation and annihilation operators, $n_{i, \sigma}=c_{i, \sigma}^{\dagger}c^{•}_{i,\sigma}$ is the number of electrons on site \textit{i} with spin $\sigma$, $t_{\pi}$ and \textit{U} are the nearest-neighbour hopping parameter and the on-site Coulomb repulsion parameter between two electrons with opposite spins, respectively.  We recall that within this Hamiltonian the on-site Coulomb interaction $U$ must be considered an effective parameter, whose value for half-filled systems with non-small values of the interactions can be taken equivalent to $U^{eff}= U - V_{1}$, where $V_{1}$ is the first nearest neighbour Coulomb interaction in the extended Hubbard models \cite{Mazumdar2012,Alfonsi2010}. Thus the (effective) $U$ parameter takes implicitly into account also longer range Coulomb interactions in the limit of static screening, as discussed by Wehling \textit{et al.} \cite{Bluegel2011}.
\\
The one-photon optical spectral function of the AGNR is calculated according to the Lehmann representation
\begin{equation}\label{spectralfun}
I(E)= \sum_{m} \vert\langle\psi_{m}\vert v^{2D}_{\alpha}\vert \psi_{GS}\rangle\vert^{2}\delta\left(E + E_{GS} - E_{m}\right)
\end{equation}
 where $E_{GS}$ is the ground-state (GS) energy of the system and $E_{m}$ the energy of any other eigenstate $\vert\psi_{m}\rangle$ obtained from exact diagonalization of the Hubbard Hamiltonian and $v^{2D}_{\alpha}$ is the velocity operator for light polarization either along $\left(\parallel\right)$ or perpendicular $\left(\perp\right)$ to the GNR edges, whose general form can be expressed concisely as
\begin{equation}\label{velocgen}
v^{2D}_{\alpha}=-\frac{i t_{\pi}}{\hbar}\sum_{\langle i,j\rangle , \sigma}\left(c^{\dagger}_{i,\sigma}c^{•}_{j,\sigma} - c^{\dagger}_{j,\sigma}c^{•}_{i,\sigma}\right)_{\alpha}
\end{equation}
The extended expressions of $v^{2D}_{\alpha=\parallel,\perp}$ are reported for completeness in the Appendix for the 4-AGNR sample system and can be straightforwardly derived for the remaining structures. 
The operator $v^{2D}_{\parallel}$ obeys the selection rule for the azimuthal band quantum number $\mu$, namely $\Delta \mu=0$ for interband transitions in AGNRs with light polarization along the nanoribbon edges \cite{Sasaki2011,Sasakiunpub}. As it occurs with zigzag SWCNTs \cite{Alfonsi2010,Alfonsi2011}, this implies that vertical transitions occur between van Hove singularities belonging to the same AGNR  band quantum number in the TB picture, as noted in  our previous work on AGNRs \cite{Alfonsi2012}. There we made the choice of considering only the case of polarization along the ribbon edges as in most literature on excitonic effects in AGNRs, since  in quasi-1D materials strong depolarization effects are believed to quench the optical absorption for perpendicular polarization \cite{Prezzi2008}. However, because of the structural anisotropy of AGNRs, we expect that the optical response of these systems should exhibit anisotropic features which could in principle be detected in polarized absorption measurements, also observed by Gundra and Shukla \cite{Shukla2011}. In order to test this hypothesis, we determined the optically allowed transitions also for the $v^{2D}_{\perp}$ operator, which takes into account vertical transitions with non-conserved band quantum number, that is with $\Delta \mu = \pm 1, \pm 3, \ldots $\\
The non-zero matrix elements of $v^{2D}_{\alpha}$ between each generic eigenstate $\vert m\rangle$ and the one-photon allowed eigenstates previously determined through Eq.\ref{spectralfun} allow us to recognize the two-photon allowed states in the obtained eigenset.\\
We consider a half-filled system with \textit{n} electrons distributed over $n = 2N$ sites and total spin quantum number $S_{z} = 0$. Thus the size of the basis set and the dimension of the matrix to be diagonalized is $D = \left[n!/(n_{\uparrow}! n_{\downarrow}!)\right]^{2} $, where $n_{\uparrow}$ and $n_{\downarrow}$ are the numbers
of spin-up and spin-down electrons, respectively, with $n_{\uparrow} = n_{\downarrow}= N$.
Calculations were performed for several values of the $U/t$ correlation coupling
strength. An additional set of calculations was performed for non-zero $S_{z}$, namely
$S_{z} = (n_{\uparrow} - n_{\downarrow})/2$, in order to identify among the obtained
$S_{z}= 0$ eigenstates the corresponding spin multiplicity. In this way, only those $S_{z} = 0$ eigenstates which also appear in the $S_{z} = 1$ set but not in the $S_{z} = 2$ set are classified as triplet excitations, whereas the remaining $S_{z} = 0$ eigenstates, which do not belong to the $S_{z} = 1$ set at all, are classified as singlets.
We verified that in the low-energy spectrum, it is sufficient to perform ED up to the $S_{z} = 2$ quantum number, since states belonging to the $S_{z} = 3$ set do occur well above the energies of the states of interest for the low-energy optical properties. 
In order to perform ED for the Hubbard model, we adopted an iterative diagonalization scheme based on the Lanczos algorithm, as implemented in the ALPS libraries \cite{ALPS2}, and additional matrix-free strategy combined with shared-memory parallelization on multicore nodes \cite{Schnack2008} in order to speed-up the diagonalization of the larger systems with 12 and 14 sites mimicking the 6- and 7-AGNRs, respectively.

\section{Results and discussion}
\begin{figure}[htbp!]
\epsfxsize=8cm
\epsfysize=7cm
\centering
\includegraphics[width=0.8\linewidth]{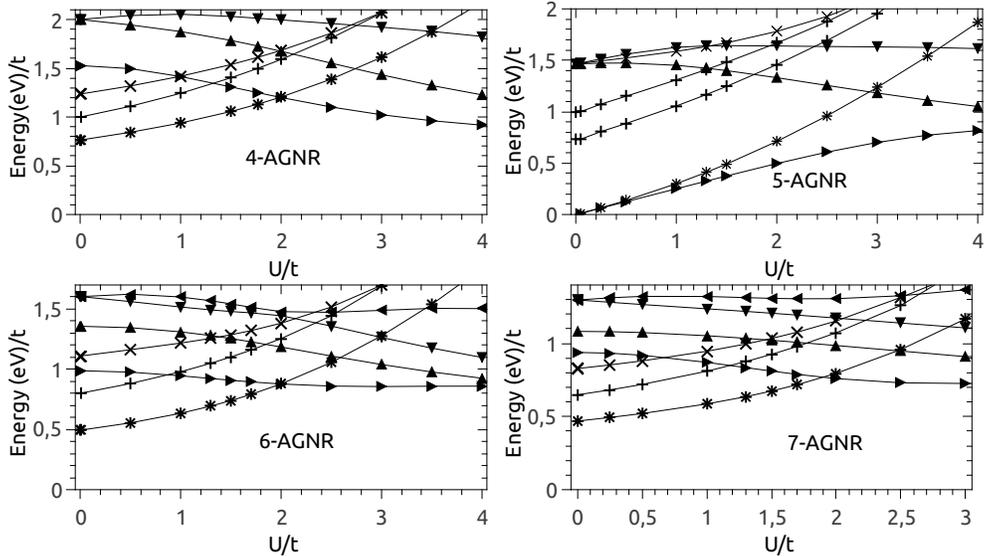}
\caption{Transition energies vs $U/t$ for the considered \textit{N}-AGNRs with $N=4,5,6,7$. Symbol legends: $\left(\ast\right)$ one-photon first bright state E11 obtained from $v^{2D}_{\parallel}$; $\left(+ \right)$ one-photon bright state(s) obtained from $v^{2D}_{\perp}$; $\left(\times\right)$ one-photon second bright state E22 from $v^{2D}_{\parallel}$; two-photon states are denoted by triangles.}
\label{transvsUt}
\end{figure}

\begin{figure}[htbp!]
\epsfxsize=8cm
\epsfysize=7cm
\centering
\includegraphics[width=0.8\linewidth]{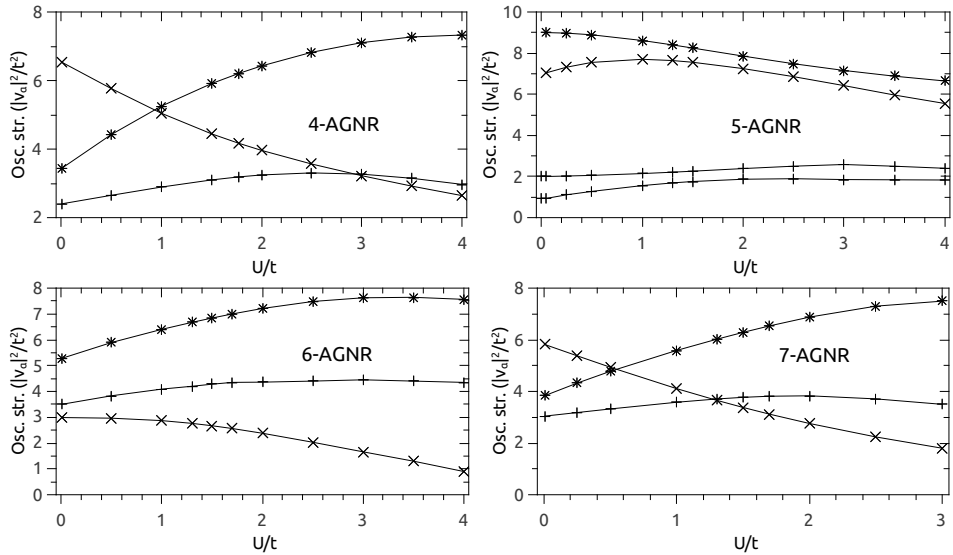}
\caption{Oscillator strengths vs $U/t$ for the considered \textit{N}-AGNRs with $N=4,5,6,7$. Symbol legends: $\left(\ast\right)$ one-photon first bright state E11 obtained from $v^{2D}_{\parallel}$; $\left(+ \right)$ one-photon bright state(s) obtained from $v^{2D}_{\perp}$; $\left(\times\right)$ one-photon second bright state E22 from $v^{2D}_{\parallel}$.}
\label{optvsUt}
\end{figure}

In Fig.\ref{transvsUt} a general overview of the one- and two-photon allowed transitions is given for either longitudinally and transversely polarized photons in AGNRs with $N=4 \div 7$ in the correlation regime $0 < U/t < 4$. One-photon allowed transitions obtained from the perpendicularly-polarized velocity operator $v^{2D}_{\perp}$ are denoted by the $\left(+\right)$ symbol, whereas the ES11 and ES22 obtained from the $v^{2D}_{\parallel}$ operator are denoted by the $\left(\ast\right)$ and $\left(\times\right)$ symbols, respectively. We find that both components of the velocity operator $v^{2D}$ couple  the obtained one-photon allowed states to the same set of two-photon allowed transitions.
For the case $U/t=0$ the TB results were verified for the interband optical transition energies and the metallicity condition, $N=3p+2$, with \textit{p}  an integer, for 5-AGNR. For the remaining ribbon families with $N=3p$ (6-AGNR) and $N=3p+1$ (4- and 7-AGNR)   we verified the semiconducting behaviour. However, we recall that when
electronic correlations are taken into account $U/t > 0$ the metallic behaviour of the 5-AGNR is no longer observed because of the lifting of the K-point degeneracy and all the three AGNRs families display semiconducting behaviour.
\\
As a feature common to all the three AGNR families, the bright states given by $v^{2D}_{\perp}$ are always found between the ES11 and ES22 transitions obtained from $v^{2D}_{\parallel}$.
Using the values $t_{\pi}=\unit[2.6\div 2.8]{eV}$ for the TB hopping parameter in graphene-based materials \cite{Cini2008,Bluegel2011}, the magnitudes of the one-photon transition energies are of the same order of those  previously computed by GW Bethe-Salpeter methods \cite{Louie2007, Prezzi2008} or by the PPP Hamiltonian \cite{Shukla2011,Mazumdar2012} for the corresponding families.\\
\\
Interestingly, we note that $v^{2D}_{\perp}$ activates only one transition in the $N=3p,3p+1$ families, whereas two one-photon allowed transitions are found for perpendicular light polarization for the $N=3p+2$ family (5-AGNR). This finding again confirms our previous result that the opto-electronic properties of AGNRs are family-dependent, as also observed by Prezzi et al. \cite{Prezzi2008}, although the behaviour of the $N=3p+2$ family is significantly different from those of the $N=3p,3p+1$ systems \cite{Alfonsi2012}.\\
Since graphene nanoribbons are in the intermediate-low correlation regime $1 \leq U/t \leq 2$ \cite{Yazyev2010}, at least one two-photon allowed transition is always found between ES11 and ES22 for all the three AGNR families, while the one-photon transition given by $v^{2D}_{\perp}$ can be found either below $\left( U/t \approx 1\right)$ or above $\left( U/t \approx 2\right)$ the two-photon allowed state.\\
In Fig. \ref{optvsUt} the magnitude trends of the optical oscillator strengths for the considered one-photon allowed transitions are reported as a function of the correlation strength $U/t$. \\
\\
Again for the $N=3p+2$ family, we note a strikingly different behaviour than in the other systems, since the $t^2$-normalized optical oscillator strengths of the $v^{2D}_{\perp}$ transitions are quite low (about 4 times lower than the ES11 or ES22 oscillator strengths), whereas in the $N=3p,3p+1$ families the reported values for the corresponding transition are comparable or even higher (6-AGNR) than the ES22 transition.
Since the $v^{2D}_{\perp}$ transitions are well separated in energy from those obtained with longitudinally polarized photons, these results predict that anisotropic optical features are likely to be observed more distinctly in AGNRs belonging to the $N=3p,3p+1$ families.\\
\begin{figure}[htbp!]
\epsfxsize=8cm
\epsfysize=7cm
\centering
\includegraphics[width=0.9\linewidth]{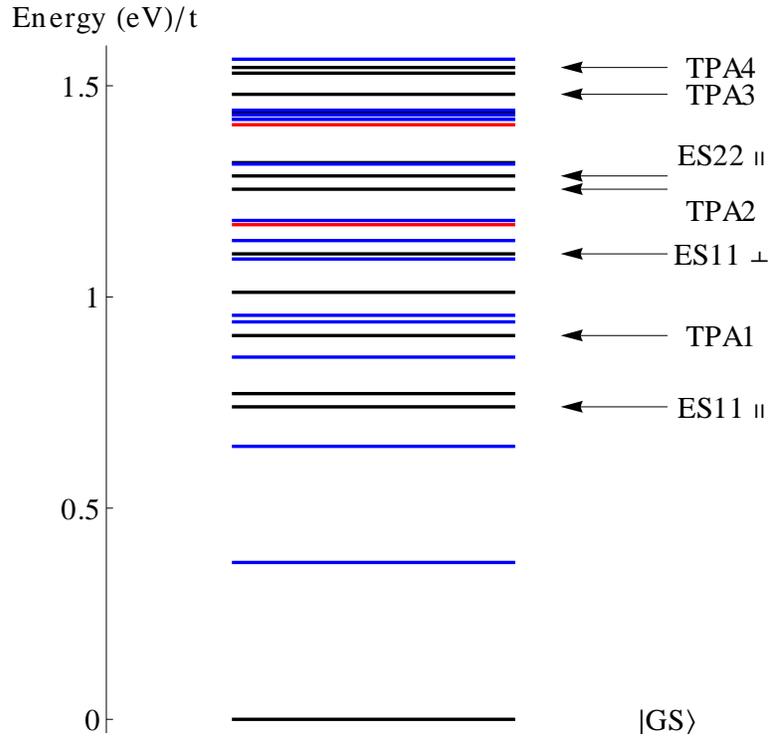}
\caption{Exciton level scheme of 6-AGNR ($N=3p$) obtained from ED  with $U/t=1.5$. Level legends: singlets (black), triplets (blue), quintuplets (red). $\mathrm{ES11, \, ES22}_{\parallel}$ are the one-photon  allowed transitions obtained from $v^{2D}_{\parallel}$, $\mathrm{ES11}_{\perp}$ from $v^{2D}_{\perp}$, TPA1-4 are the two-photon allowed states.}
\label{levels6agnr}
\end{figure}
\\
In Fig. \ref{levels6agnr} we report the low-energy excited-state spectrum of 6-AGNR with an explicit characterization of the spin multiplicity of the optically inactive (dark) exciton states alongside the allowed one- and two-photon transitions for both polarization directions. The 6-AGNR was chosen since according to Fig. \ref{optvsUt} this system displays high values of the ES11 oscillator strengths for both polarization components in the considered correlation regime. Anyway, a very similar fine structure of the exciton levels can also be obtained for 7-AGNR. 
We observe the presence of deep-lying triplet excited states above the ground state, as reported in SCI calculations by Dutta et al.\cite{Pati2008}. Moreover a dark singlet is found $\unit[81]{meV}$ above $\mathrm{ES11}_{\parallel}$ for $t=\unit[2.6]{eV}$, whereas two triplets are found $\unit[82]{meV}$ above and $\unit[31]{meV}$ below $\mathrm{ES11}_{\perp}$, respectively. As discussed in the case of SWCNTs \cite{Alfonsi2011}, these states can act as a population sink for the nearby optically-allowed exciton through a bottleneck mechanism. In particular, the spin-orbit coupling due to impurities or defects is expected to enhance the quenching of the radiative transition by trapping the radiation in triplet dark states, so that the luminescence quantum yield from the bright state would be decreased. 

\section{Conclusions}
In summary, we have investigated the excitonic structure of pristine small-width AGNRs by ED of the Hubbard model for several two-leg ladder models mimicking ribbons with $4 \leq N \leq 7$ dimer lines and for both components of photon polarizations, either along and perpendicular to the ribbon edges. By this technique we are able to investigate the effect of electronic correlations involved in direct interband optical transitions by a full-many body approach for several values of the Hubbard correlation parameter $U/t$. The obtained results allow us to track down the nanoribbon family dependence of the  anisotropic features in the optical properties of these systems, which could be detected in polarized absorption measurements.

\begin{acknowledgments}
Work supported by University of Padova under grant no. CPDR091818. The CINECA award under the ISCRA initiative HP10C4ZPOY (2011) is gratefully acknowledged for the availability of high performance computing resources and support.
\end{acknowledgments}

\appendix
\section{Velocity operators for interband optical transitions in armchair GNRs}
\begin{figure}[htbp!]
\epsfxsize=8cm
\epsfysize=7cm
\centering
\includegraphics[width=0.4\linewidth]{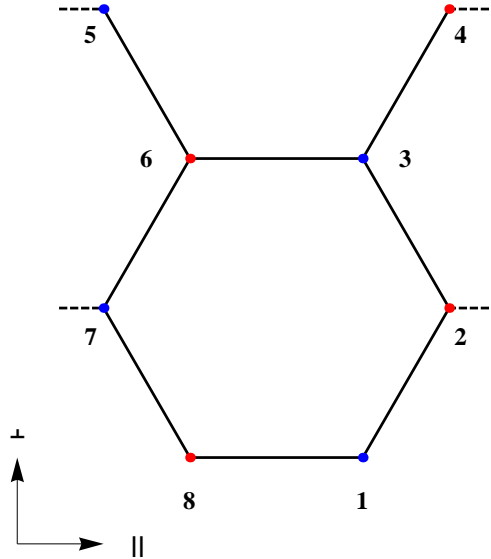}
\caption{(color online) Geometrical structure of the 4-AGNR.}
\label{latagnr}
\end{figure}

In the following we report the explicit second-quantization expressions for the components of the velocity operator $v^{2D}_{\alpha=\parallel,\perp}$ along and perpendicular to the edges of the 4-AGNR. The site indexing of the creation/annihilation operators refers to the atomic labels reported in the related Fig. \ref{latagnr}.
\begin{eqnarray}\nonumber
v_{\parallel}^{N=4} = - \frac{i t_{\pi}}{\hbar}\sum_{\sigma}\left[ -\left(c^{\dagger}_{1,\sigma}c^{•}_{8,\sigma} + c^{\dagger}_{3,\sigma}c^{•}_{6,\sigma} - \mathrm{h.\,c.}\right) + \right. \qquad\\
+ \frac{1}{2} \left(c^{\dagger}_{1,\sigma}c^{•}_{2,\sigma} + c^{\dagger}_{3,\sigma}c^{•}_{4,\sigma} + c^{\dagger}_{7,\sigma}c^{•}_{6,\sigma} - \mathrm{h.\,c.}\right) + \qquad\\
\nonumber
+\left(c^{\dagger}_{2,\sigma}c^{•}_{7,\sigma} + c^{\dagger}_{4,\sigma}c^{•}_{5,\sigma} - \mathrm{h.\,c.}\right) - \left. \frac{1}{2}\left(c^{\dagger}_{2,\sigma}c^{•}_{3,\sigma} + c^{\dagger}_{8,\sigma}c^{•}_{7,\sigma} + c^{\dagger}_{6,\sigma}c^{•}_{5,\sigma} - \mathrm{h.\,c.}\right) \right]
\label{vel4agnrparal}
\end{eqnarray}
\begin{eqnarray}
v_{\perp}^{N=4} = -\frac{i t_{\pi}}{\hbar}\frac{\sqrt{3}}{2}\sum_{\sigma}\left[ c^{\dagger}_{1,\sigma}c^{•}_{2,\sigma} + c^{\dagger}_{2,\sigma}c^{•}_{3,\sigma} + c^{\dagger}_{3,\sigma}c^{•}_{4,\sigma} + c^{\dagger}_{6,\sigma}c^{•}_{5,\sigma} + \quad \right. \\
\left.
\nonumber
+ \, c^{\dagger}_{7,\sigma}c^{•}_{6,\sigma} + c^{\dagger}_{8,\sigma}c^{•}_{7,\sigma} - \mathrm{h.\,c.} \right]
\label{vel4agnrperp}
\end{eqnarray}
\\

\end{document}